\documentclass[12pt,preprint]{aastex}
\usepackage{emulateapj5}

\def\be{\begin{equation}}
\def\ee{\end{equation}}
\def\ba{\begin{eqnarray}}
\def\ea{\end{eqnarray}}

\def\bB{{\bf B}}

\def\go{\mathrel{\raise.3ex\hbox{$>$}\mkern-14mu
             \lower0.6ex\hbox{$\sim$}}}
\def\lo{\mathrel{\raise.3ex\hbox{$<$}\mkern-14mu
             \lower0.6ex\hbox{$\sim$}}}

\newenvironment{inlinefigure}{
\medskip
\def\@captype{figure}
\noindent\begin{minipage}{0.999\linewidth}\begin{center}}
{\end{center}\end{minipage}\medskip}

\begin{document}
\title{The Hidden Magnetic Field of The Young Neutron Star in Kesteven 79}
\author{Natalia Shabaltas\altaffilmark1 and Dong Lai\altaffilmark1}
\altaffiltext1{Center for Space Research, Department of Astronomy, 
Cornell University, Ithaca, NY 14853, USA}

\begin{abstract}
Recent observations of the central compact object in the Kesteven 79
supernova remnant show that this neutron star (NS) has a weak
dipole magnetic field (a few$\times 10^{10}$G) but an anomalously
large ($\sim 64\%$) pulse fraction in its surface X-ray emission.  We
explore the idea that a substantial sub-surface magnetic field exists
in the NS crust, which produces diffuse hot spots on the stellar
surface due to anisotropic heat conduction, and gives rise to the
observed X-ray pulsation.  We develop a general-purpose method, termed
``Temperature Template with Full Transport'' (TTFT), that computes the
synthetic pulse profile of surface X-ray emission from NSs
with arbitrary magnetic field and surface temperature distributions,
taking into account magnetic atmosphere opacities, beam pattern,
vacuum polarization, and gravitational light bending.  We show that 
a crustal toroidal magnetic field of order a few$\times 10^{14}$G or 
higher, varying smoothly across the crust, can produce sufficiently 
distinct surface hot spots to generate the observed pulse
fraction in the Kes 79 NS. This result suggests that substantial
sub-surface magnetic fields, much stronger than the ``visible'' dipole
fields, may be buried in the crusts of some young NSs, and such hidden
magnetic fields can play an important role in their observational
manifestations. The general TTFT tool we have developed can also be used
for studying radiation from other magnetic NSs.
\end{abstract}


\section{Introduction}

Magnetic field is perhaps the single most important quantity that
determines the observational manifestations of isolated
neutron stars (NSs; see, e.g., Turolla 2009; Kaspi 2010).
Most of our knowledge of NS magnetic fields is based on the
measurement of the rotation period $P$ and period derivative
$\dot{P}$, which, assuming that NS spin-down is due to angular
momentum loss through magnetic dipole radiation or magnetic wind, 
yields estimates for the dipolar magnetic field strength of many NSs.
Several NSs are observed to have spectral lines which, when interpreted as
electron/ion cyclotron lines or atomic lines, yield a value for the
\textit{surface} magnetic field strength (e.g., Ho \& Lai 2004; van
Kerkwijk \& Kaplan 2007; Suleimanov et al. 2010). 
Neither of these measurements sheds any light on the internal magnetic field
configuration. 
Several recent observations suggest that the internal NS magnetic field must
be more complex than a simple dipole. For example, the pulse profiles
of X-ray Dim Isolated Neutron Stars may be explained by the
addition of higher order multipoles (Zane \& Turolla 2006).
Also, magnetars are thought to require large toroidal
fields to fuel their bursting activity (e.g., Thompson \& Duncan
2001). The most striking example of a NS undergoing magnetar-like
bursts is SGR 0418+5729, whose observed dipole
field ($\lo 8\times 10^{12}$G) is much too low compared with typical
magnetars, and therefore a significant hidden field must be present to
power it (Rea et al. 2010; Turolla et al.~2011).

The central compact object (CCO) PSR J1852+0040 in the Kesteven 79
supernova remnant is another source with unusual
properties (Halpern \& Gotthelf 2010). Specifically, while its dipole
surface magnetic field is inferred via $P$-$\dot{P}$ measurement 
($P=0.105$~s, $\dot P=8.7\times 10^{-18}~\mathrm{s~s^{-1}}$) to be
only $\sim 3.1\times10^{10}$~G, its pulse profile presents with an
unusually high pulse fraction ($64\pm 2\%$), which implies large
temperature anisotropy (concentrated hot spots) on its surface.
Such hot spots are unattainable by anisotropic heat conduction at the
observed dipole field strength. Also, since the spindown luminosity is
much smaller than the X-ray luminosity, they cannot be produced by
heating from magnetospheric activities.
In this paper we explore a more complex internal field
configuration, adding a large crustal toroidal field aligned about the
magnetic dipole axis to suppress heat conduction everywhere except
toward the magnetic poles of the NS. 
While the idea of using crustal field to suppress heat conduction is
not new (e.g., Geppert et al. 2004, 2006),
our goal is to use the observed X-ray light curve to quantitatively 
constrain the magnitude and shape of this hidden toroidal field.

To quantitatively model the X-ray light curve of a magnetic NS, it is
necessary to add up emissions from all surface patches of the star, taking
account of beaming/anisotropy due to magnetic fields and light bending
due to general relativity (Zane \& Turolla 2006). While this is conceptually straightforward,
it is not easy to implement in practice, as atmosphere models for many
surface patches (each with different $T_{\rm eff}$ and ${\bf B}$) are
needed. Therefore, in Section 2 we develop a practical method,
termed ``Temperature Template with Full Transport'' (TTFT), for 
computing radiation intensity from different surface patches.
This method captures the essential features of radiation from 
magnetic atmospheres, but does not require full self-consistent atmosphere 
modeling for each surface patch. The TTFT method allows us to 
compute NS light curves with arbitrary surface magnetic field
and temperature distributions.
In Section 3, we study various crustal magnetic field configurations, present the resulting
surface effective temperature distributions, and apply the method of
Section 2 to compute the pulse profiles and reproduce the
observed high pulse fraction of the CCO in Kes 79. We discuss the results in 
Section 4, and conclude with Section 5.

\section{Light Curve Calculation and the TTFT Method}

In this section we outline our method for computing X-ray light curves
for NSs with arbitrary surface magnetic field and temperature
distributions. Although our application focuses on the CCO in Kes 79
(Section 3), the method presented here is general and can be applied to
other types of magnetic NSs with atmospheres.
\footnote{We do not consider the extreme case of high magnetic field at low temperature, for which the NS surface may be in condensed form (e.g., Lai \& Salpeter 1997; Medin \& Lai 2007).}
 
\subsection{Light Curve Calculation}

The observed specific radiation flux $dF^o_\nu$ due to an infinitesimal
patch (area $dA$) on the NS surface is given by
\begin{equation}
dF^o_\nu = I^o_\nu d\Omega^o = {g\over{D^2}}I^e_\nu(\textbf{k})\cos\alpha\left|d\cos\alpha\over{d\mu}\right|dA,
\label{eq:dFonu1}
\end{equation}
where $I^o_\nu$ is the observed specific intensity; $d\Omega^o$ is the
observed angular size; $g$ is the gravitational redshift factor given by
$g^2\equiv\left(1-2GM/Rc^2\right)$; $M$, $R$, and $D$ are the mass,
radius, and distance to the NS; $I^e_\nu$ is the local specific intensity
(as emitted at the surface patch of the NS); $\textbf{k}$ is the direction of
photon propagation at the surface; 
$\alpha\equiv\cos^{-1}(\textbf{k}\cdot\hat\textbf{r})$ is the angle
between $\textbf{k}$ and the local radial vector; and $\mu$ is the cosine
of the polar angle (with respect to the observation axis) of the patch.  Due to gravitational
light bending, $\cos\alpha$ is a non-trivial function of
$\mu$. Following Beloborodov (2002), we make an approximation to this
function, and calculate the observed specific flux as
\begin{equation}
F^o_\nu\simeq\int {g^3\over{D^2}}I^e_\nu({\bf k})\left[1-g^2\left(1-\mu\right)\right]dA,
\label{eq:dFonu3}
\end{equation}
where the integration is over the entire observable area (including light
bending) of the NS surface. Thus, given the knowledge of local
radiative intensity [$I^e_\nu(\textbf{k})$] of all surface patches,
combined with the photon trajectory at different rotational phase
(time) [$\mu(t) \Leftrightarrow \alpha(t)$], the resultant observed
specific flux can be calculated.

\subsection{Radiative Transfer in Magnetic Neutron Star Atmospheres}

The full computation of $I^e_\nu(\textbf{k})$ requires construction of
atmospheric models in radiative equilibrium for various surface
magnetic fields $\bB$ (magnitude and direction), effective
temperatures $T_{\mathrm{eff}}$, and compositions (e.g., 
Shibanov et al.~1992; Pavlov et~al.~1995; Zane et al.~2001; Ho \& Lai 2001,~2003,~2004; 
Potekhin et~al.~2004; van Adelsberg \& Lai 2006; Suleimanov et~al.~2010).  
This is currently not practical for exploration of a wide range of
model parameters, since quantitative calculation of the light curve
demands adding up many different surface patches (rather than one or
two hot spots; cf.~Ho 2007). 
Therefore we develop an approximate method (TTFT)
(see Section 2.3) that allows us to obtain $I^e_\nu(\textbf{k})$
efficiently. Here, we describe our full radiative transfer calculation
in NS atmospheres.

In the atmospheres of magnetized NSs, photon polarization plays an
important role in determining $I^e_\nu(\textbf{k})$.  For example, at
photon frequencies smaller than the electron cyclotron frequency, the
mean free path for ordinary (O) mode photons is much smaller than that
of the extraordinary (X) mode photons, and therefore, the X-mode
photons that contribute to the emergent flux are typically produced in
the deeper, hotter layers of the atmosphere. At higher field strengths
($\gtrsim8\times10^{13}$G), vacuum polarization will also have a
strong effect on the radiation spectrum (Lai \& Ho 2002,~2003). While
rigorous treatment of this effect is available (van Adelsberg \& Lai
2006), we describe an approximate method below.

For a photon of a given energy $E$ propagating along $\textbf{k}$ in
the NS atmosphere, vacuum resonance occurs at
\begin{equation}
\rho_V=0.96 Y_e^{-1} E_1^2 B_{14}^2 f_B^{-2}~\mathrm{g~cm^{-3}},
\label{eq:rhov}
\end{equation}
where $Y_e$ is the electron fraction, $E_1=E/(1~\mathrm{keV})$,
$B_{14}=B/(10^{14}~\mathrm{G})$, and $f_B$ is a slowly varying
function of $B$, of order unity. At resonance, the X and O
polarization modes become degenerate, and therefore the photon may be
converted from one mode to another when it traverses the resonance. We
define two more effective polarizations modes: + and --. These modes
act as X (O) polarization modes before the resonance and as O (X)
modes after. Lai \& Ho (2002) showed that photons satisfying the
adiabatic condition, $E \gtrsim E_{\mathrm{ad}}$, act as + and --
modes. Here
\begin{equation}
E_{\mathrm{ad}}=2.52\Bigl[f_B \tan\theta_{kB}\left|1-(E_{Bi}/E)^2\right|
\Bigr]^{2/3}\left({1~\mathrm{cm}\over H_\rho}\right)^{1/3}\mathrm{keV},
\label{eq:ead}
\end{equation}
where $\theta_{kB}\equiv\cos^{-1}\textbf{k}\cdot\hat{\textbf{B}}$,
$E_{Bi}=0.63Y_e B_{14}~\mathrm{keV}$ is the ion cyclotron energy, and
$H_\rho\equiv|ds/d\ln \rho|$ is the density scale height along the
ray. More generally, a photon with energy $E\sim E_{\mathrm{ad}}$ will
have a nonadiabatic `jump' probability of
\begin{equation}
P_{\mathrm{jump}}=\exp\left[-{\pi\over 2}\left(E/E_{\mathrm{ad}}\right)^3\right].
\label{eq:pjump}
\end{equation}
If the `jump' probability is low, most outward diffusing photons will
evolve through the resonance adiabatically as + or -- modes,
undergoing mode conversion. On the other hand, if the `jump'
probability is high, X-mode (O-mode) photons will remain X-mode
(O-mode) photons after traversing the resonance. We therefore elect to
treat the X, O, +, and -- photon modes as distinct carriers whose
relative abundances are determined by $P_{\mathrm{jump}}$:
\begin{equation}
I^e_\nu=P_{\mathrm{jump}}(I^X_\nu+I^O_\nu)+(1-P_{\mathrm{jump}})(I^+_\nu+I^-_\nu).
\label{eq:inucarriers}
\end{equation} 
Note that, for the application to the Kes 79 CCO (Section 3), 
at fields $B\sim 10^{10}-10^{11}$G and energies $E\sim1-5$keV, $\rho_V$ 
is smaller than the photosphere densities, and therefore vacuum
polarization does not play an important role. We therefore treat all
photons as ``jump'' photons ($P_{\mathrm{jump}}=1$), and do not compute intensities for the +
and -- modes.

In general, to compute the individual specific intensities for each carrier, we integrate the standard radiative transfer equation, $dI^i_{\nu}/d\tau^i_{\nu}=B_\nu-I^i_\nu$. However, for the present purposes 
it is sufficient to use the Eddington-Barbier relation:

\begin{equation}
I^i_\nu(\textbf{k})\approx {1\over 2}B_\nu\left[T(\tau^i_\nu(\textbf{k})=2/3)\right].
\label{eq:eddington}
\end{equation}

\begin{inlinefigure}
\scalebox{0.55}{\rotatebox{0}{\includegraphics{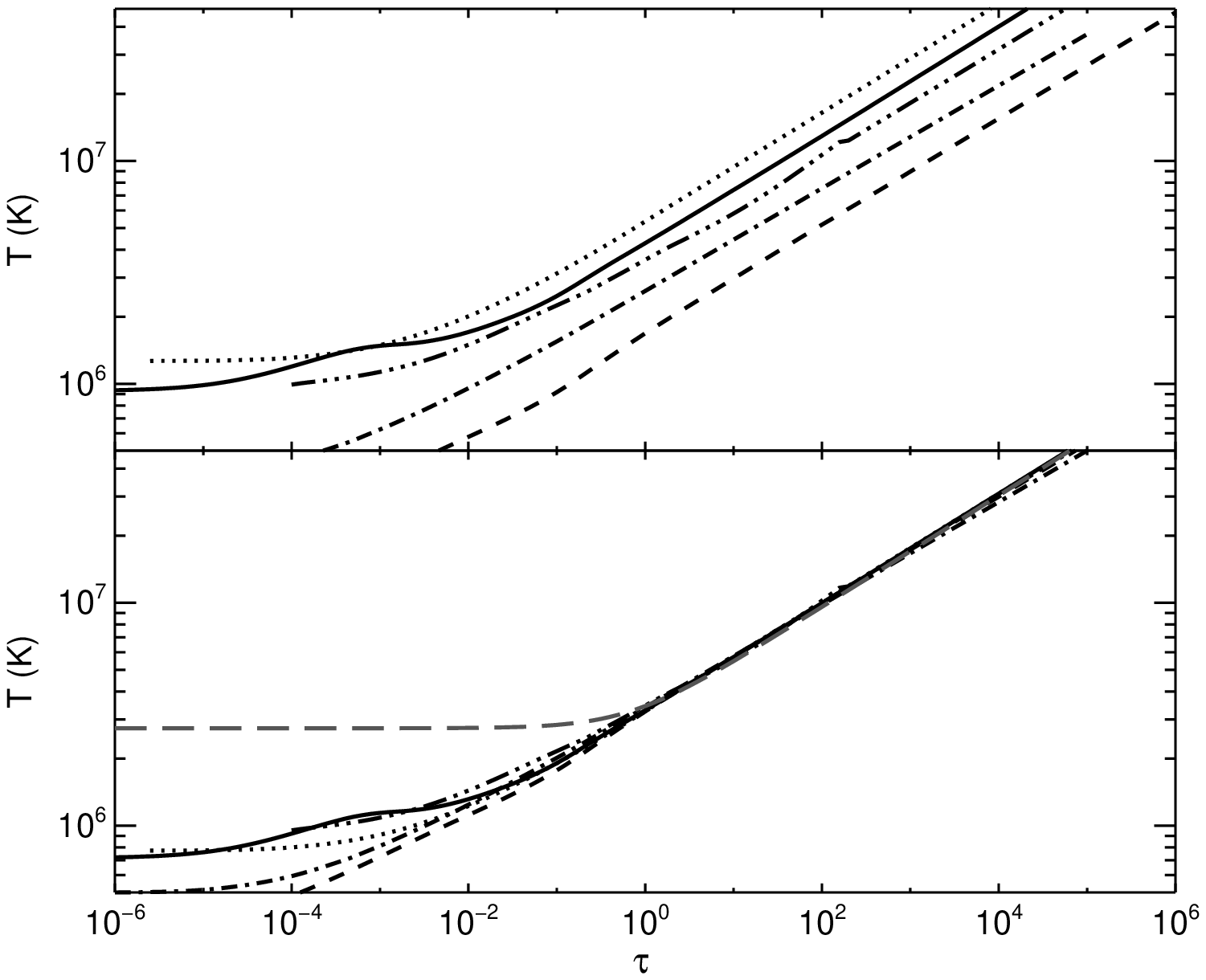}}}
\figcaption{{\it Top panel:} NS atmosphere temperature
  profiles vs Thomson depth from several self-consistent atmosphere models.
  Solid line -- $B=6.2\times10^{10}$~G, $T_{\mathrm{eff}}=4\times10^{6}$~K; 
  dashed line -- $B=6.2\times10^{10}$~G, $T_{\mathrm{eff}}=1\times10^{6}$~K;
  dotted line -- $B=0$, $T_{\mathrm{eff}}=5\times10^{6}$~K;
  dashed-dotted line -- $B=0$, $T_{\mathrm{eff}}=1.6\times10^{6}$~K;
  dashed-triple-dotted line -- $B=2\times10^{12}$~G, $T_{\mathrm{eff}}
  =5\times10^{6}$~K. 
{\it Bottom panel:} The same profiles rescaled such that
  $T_{\mathrm{eff}}=3.8\times10^6$~K at a normal magnetic field with strength
  $B=8.8\times10^{10}$~G using the TTFT method. In addition, we include the Thomson grey opacity temperature profile $T^4\sim(\tau+2/3)$(long dashed), similarly rescaled.}
\end{inlinefigure}

That is, the emergent intensity (toward $\textbf{k}$) due to carrier
$i$ is equal to the Planck function evaluated at the depth at which
the optical depth $\tau^i_\nu$ for that carrier (along $\textbf{k}$,
measured from the outside in) is equal to $2/3$. The factor of $1/2$
arises because the photon number distribution is split among two
carriers at all times. Thus, to calculate the total emergent specific
intensity, only a few pieces of information are required. These are:
the direction and magnitude of the magnetic field, the opacities for
each carrier, and the atmospheric temperature and density profiles
[$T(\tau)$ and $\rho(\tau)$, where $\tau=\kappa_T y$ (with
$\kappa_T=0.4$~cm$^2$g$^{-1}$) is the (zero-field) Thomson depth, and $y$ is the
column density].

We use free-free absorption opacities for each carrier, including the
contribution of electron-ion Coulomb collisions, as described by Lai
\& Ho (2003).  This opacity depends strongly on
$\textbf{k}\cdot\hat{\textbf{B}}$, as well as on density and
temperature.

In addition, at the magnetic field strengths
we are considering, the opacity is further complicated by oscillations
in the magnetic Gaunt factor (Potekhin 2010). However, the energies we consider have values of $\sim 1.5-5$ times the electron cyclotron energy and therefore these oscillations are not very strong. Furthermore, while the oscillations may affect the specific flux emitted at certain frequencies, we do not expect them to affect pulse shape. We therefore do not include this complication in our calculation, instead choosing to use the non-magnetic Gaunt factor, which captures the trend about which the full magnetic Gaunt factor oscillates.

The temperature profile $T(\tau)$ and density profile $\rho(\tau)$ are
related by the equation of hydrostatic equilibrium. For all practical
purposes, an ideal, non-degenerate gas equation of state can be used
with better than a few percent accuracy (e.g., Ho \& Lai 2001).

\begin{inlinefigure}
\scalebox{0.55}{\rotatebox{0}{\includegraphics{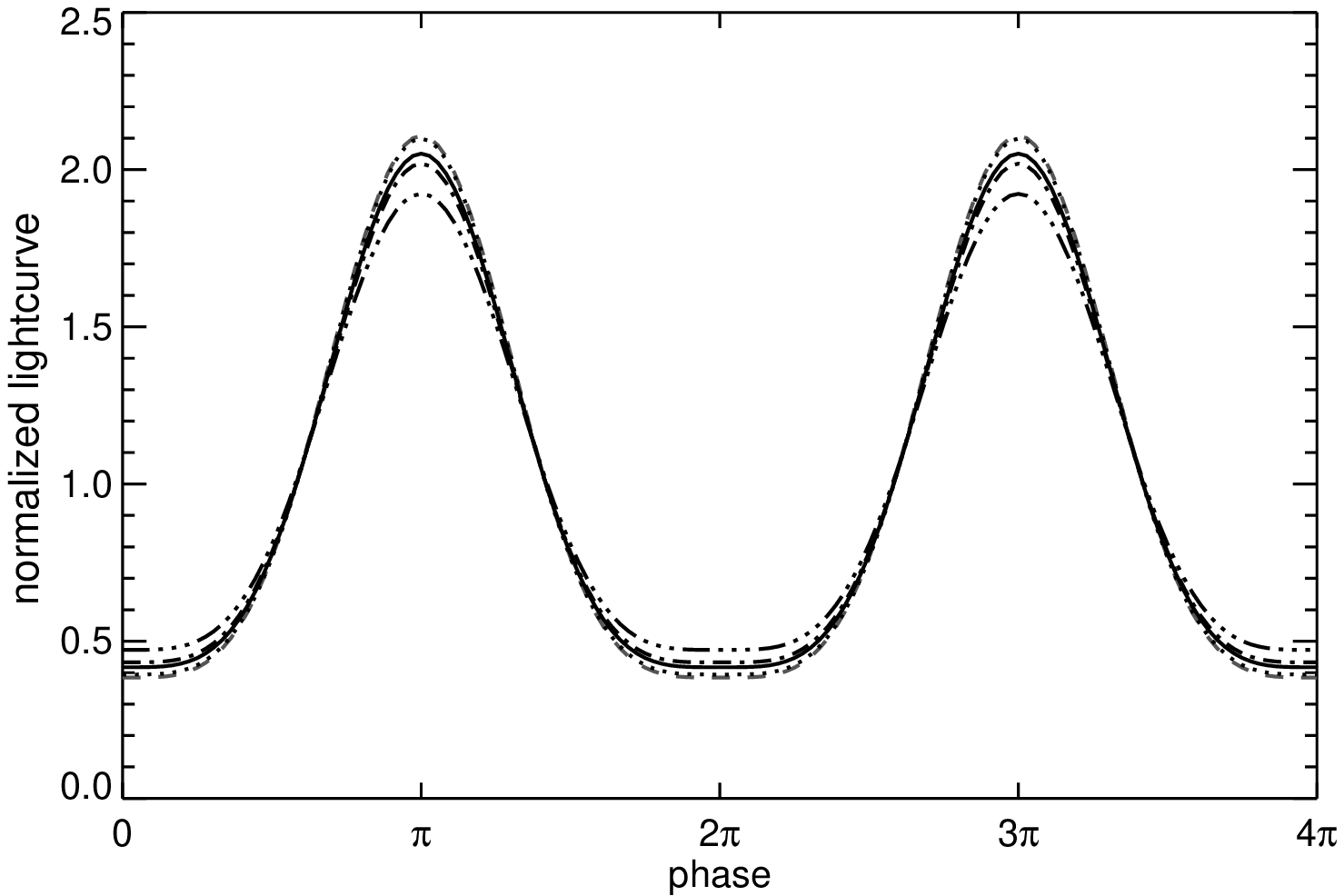}}}
\figcaption{Sample full-star light curves of NS surface
  emission at photon energies around $1.5$~keV calculated using the TTFT method. Two rotation phases are
  shown.  These light curves are computed for a specific NS surface
  temperature and magnetic field distributions and viewing geometry
  (see Section 3). The different curves correspond to the results
  obtained using different temperature profile templates as depicted
  in Figure 1.  The highest pulse fraction obtained is 62\%, and the
  lowest 53\%.}
\end{inlinefigure}

\subsection{TTFT: A Method for Temperature Profile Computation}

As an all-purpose tool, the computational procedure described in
Sections 2.1-2.2 requires the knowledge of the atmospheric temperature
profile $T(\tau)$. In theory, for a given combination of effective
temperature, magnetic field strength and direction, 
$T(\tau)$ is obtained by self-consistent atmosphere modeling
to achieve radiative equilibrium.
However, this modeling must be done for every single
patch of the NS surface, and is not practical to implement.
Here we propose the ``Temperature Template with Full Transport'' (TTFT) method
for constructing temperature profiles. While this method is 
not accurate for predicting the observed synthetic radiation spectrum, 
we show that it is reliable insofar as light curve shape 
(and therefore pulse fraction) is concerned.

The TTFT method consists of the following steps: (i) We write the temperature profile in the form 
$T(\tau)=T_0 F(\tau)$, characterized by the dimensionless
template function $F(\tau)$ and the adjustable constant $T_0$. 
The template function may be obtained from a specific NS atmosphere 
model. (ii) For each $T(\tau)$, we carry out full radiative transfer
calculation (Section 2.2) to obtain the total radiation flux from a single surface patch,
\be
{\cal F}=\int\!d\nu\!\int\!d^2{\bf k}\,\, I_\nu^e({\bf k})\,\, {\bf k}\cdot \hat{\bf r},
\label{eq:calF}\ee
where the integration $\int d^2{\bf k}$ covers the outgoing direction relative to the
surface normal vector $\hat{\bf r}$. Since ${\cal F}=\sigma T_{\rm eff}^4$, 
we then obtain a one-to-one mapping between $T_0$ 
and $T_{\rm eff}$ for the given template function $F(\tau)$.
(iii) We then adopt the {\it ansatz} that for a given surface patch with $T_{\rm eff}$, 
the atmosphere profile is $T(\tau;T_{\rm eff})=T_0(T_{\rm eff})F(\tau)$, and the
emergent radiation intensity can be computed via the method of Section 2.2.

Figure 1 demonstrates the reasoning for the validity of the TTFT
method.  The top panel presents several self-consistent temperature
profiles at different effective temperatures and field strengths
provided by Wynn Ho ($B=0$ and $B=2\times 10^{12}$~G models) and by
Valery Suleimanov ($B=6.2\times 10^{10}$~G models). We note that all
the profiles are very similar in shape, especially for $\tau\go 0.1$,
and differ mostly in overall normalization; this is true both with
respect to magnetic field strength variation and effective temperature
variation. After normalization to the same $T_{\rm eff}=3.8\times
10^6$~K and $B=8.8\times 10^{10}$~G, all the temperature profiles
become similar at $\tau\go 0.1$. Therefore, when computing the
intensity of photons that emerge from sufficient depth in the
atmosphere, the same single temperature template $F(\tau)$ may be used
for the entirety of the NS surface.

Note that it is not surprising that all the temperature profiles are
essentially identical at $\tau > 1$; indeed, in the diffusion-limited
regime that is to be expected. The less obvious and most crucial point
is that the profiles remain very similar for $0.1 < \tau < 1$, where
they also differ significantly from a grey opacity profile. In our
computations of synthetic light curves (see Figure 2), we see that
only the photons traveling nearly radially out from the star emerge
from slightly deeper than $\tau\sim 1$, and thus most of the photons
contributing to the light curve emerge from smaller Thomson
depths. Therefore, a grey opacity temperature profile could not
possibly produce the same large synthetic pulse fraction, and would
not yield a realistic representation of the emergent flux; we
\textit{must} use a self-consistent template accounting for
depth-dependent opacity.

Figure 2 shows sample X-ray light curves (normalized to a mean of
unity) of magnetic NS surface emission, computed using our TTFT method
with various template functions $F(\tau)$ depicted in Figure 1.  These
light curves are for a specific NS surface model (effective
temperature and magnetic field distributions) appropriate for the CCO
in Kes 79 (see Section 3 for details).  We see that the shapes of these
light curves are very similar, with the pulse fraction ranging from
53\% to 62\%, despite the fact that the template functions are
obtained from vastly different atmosphere models.
Note that in renormalizing the temperature profile (to obtain the 
$T_0$-$T_{\rm eff}$ mapping), we make the approximation
that the magnetic field is normal to the stellar surface, and we keep its magnitude always set to the field strength at the magnetic poles. This 
speeds up the construction of the temperature profile
(i.e., the $T_0$-$T_{\rm eff}$ mapping for a given template function)
since the integration in Equation (\ref{eq:calF}) can be done more efficiently,
and need not be repeated for every surface patch.
We have found that this approximation
introduces rather small error in the resulting 
$T_0$-$T_{\rm eff}$ relation; this is understandable  
because $T_{\rm eff}$ is determined by summing
over the emitted spectrum in all directions, and the direction of the
magnetic field is averaged over, while the radiative transfer is not too
sensitive to factor-of-two changes in field strength.
Nevertheless, we note that when computing $I_\nu^e({\bf k})$
for the light curve, the ``correct'' magnetic strength and direction must be
adopted in order to accurately account for the magnetic 
beaming effect.

\section{Constraining the Crustal Magnetic Field of the Neutron Star in Kes 79}

We now use the method described in Section 2 to model the X-ray light
curve of the NS in Kes 79, thereby constraining the magnetic field
configuration in the NS crust.

\begin{inlinefigure}
\scalebox{0.55}{\rotatebox{0}{\includegraphics{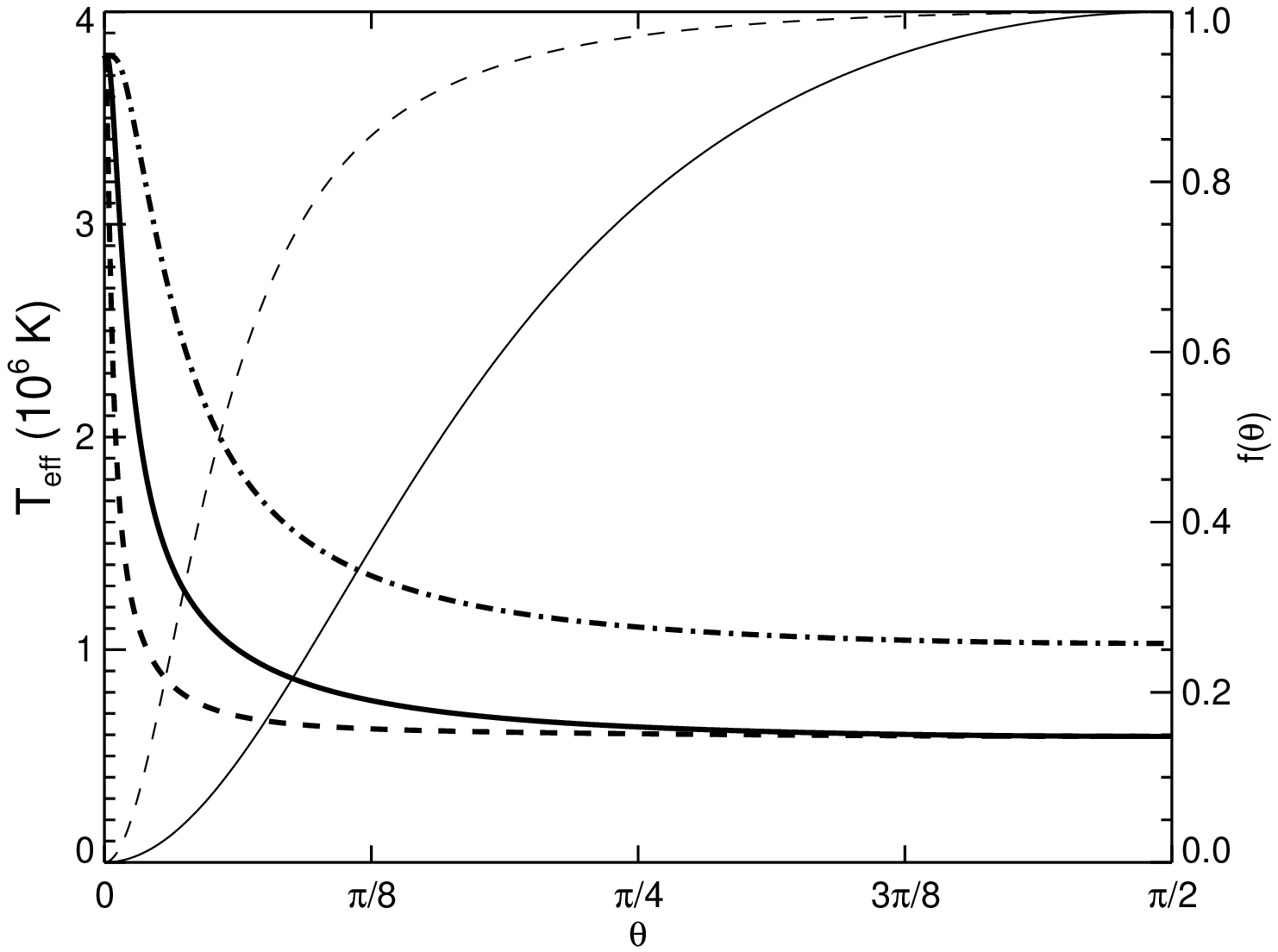}}}
\figcaption{Effective surface temperature distribution 
$T_\mathrm{eff}(\theta)$ (heavy lines) for different crustal magnetic field
configurations. The toroidal field is characterized by 
Equations (\ref{Bform})-(\ref{fform}). Solid line --
$B_0=2\times10^{14}$~G, $\theta_0=40^\circ$; dashed line --
$B_0=2\times10^{14}$~G, $\theta_0=10^\circ$; dot-dashed line --
$B_0=2\times10^{13}$~G, $\theta_0=40^\circ$.
The light lines show the corresponding function $f(\theta)$ at different 
$\theta_0$.}
\end{inlinefigure}

An internal field of strength similar to the inferred dipole field of
the Kes 79 NS would lead to a very uniform temperature distribution 
on the NS surface (e.g., Schaaf 1990; Potekhin et al.~2003).
In order to produce sufficiently concentrated hot spots,
we introduce a toroidal (with respect to the dipole axis) component to
the magnetic field, such that it is sufficiently large everywhere
except in a concentrated region around the magnetic axis. 
We adopt the following form for the toroidal field:
\begin{equation}
\label{Bform}
B_\phi(r,\theta)=B_0f(\theta)g(r),
\end{equation}
where $B_0$ is an adjustable maximum field strength
and $\theta$ is the magnetic colatitude. 

Because the NS atmosphere is likely too tenuous 
to be able to sustain a toroidal field, the toroidal field
should be contained entirely in the crust, such that the poloidal current producing it is
also contained in the crust. Thus, a realistic form of $g(r)$ should vanish at least at the NS surface,
and possibly at the crust-core interface as well. However, for simplicity, in the present calculation we treat $g(r)$ as constant.
We choose the following functional form for $f$:
\be
f(\theta)={f_0{(\sin\theta/\sin\theta_0)^2}\over{1+(\sin\theta/\sin\theta_0)^2}},
\label{fform}
\ee
where $\theta_0$ is an adjustable parameter defining how sharply the
toroidal field rises away from the magnetic poles.
The normalization constant $f_0$ is specified so that 
$\max[f(\theta)]=1$. Obviously, this functional form for $f(\theta)$ is representative;
other forms are possible. However, the detailed shape of $f(\theta)$
cannot be realistically constrained and is not important for 
our purpose. Our goal is to examine what ranges of values of $B_0$ and $\theta_0$ 
are required in order to produce sufficiently distinct surface hot spots to generate
the observed large pulse fraction in Kes 79 NS. Note that we are essentially exploring a parameter space orthogonal to the one considered by Geppert et al. (2006), who fixed $f(\theta)$ to be a simple sinusoid and explored the dependence of surface temperature and pulse shape on $g(r)$. 

\begin{inlinefigure}
\scalebox{0.55}{\rotatebox{0}{\includegraphics{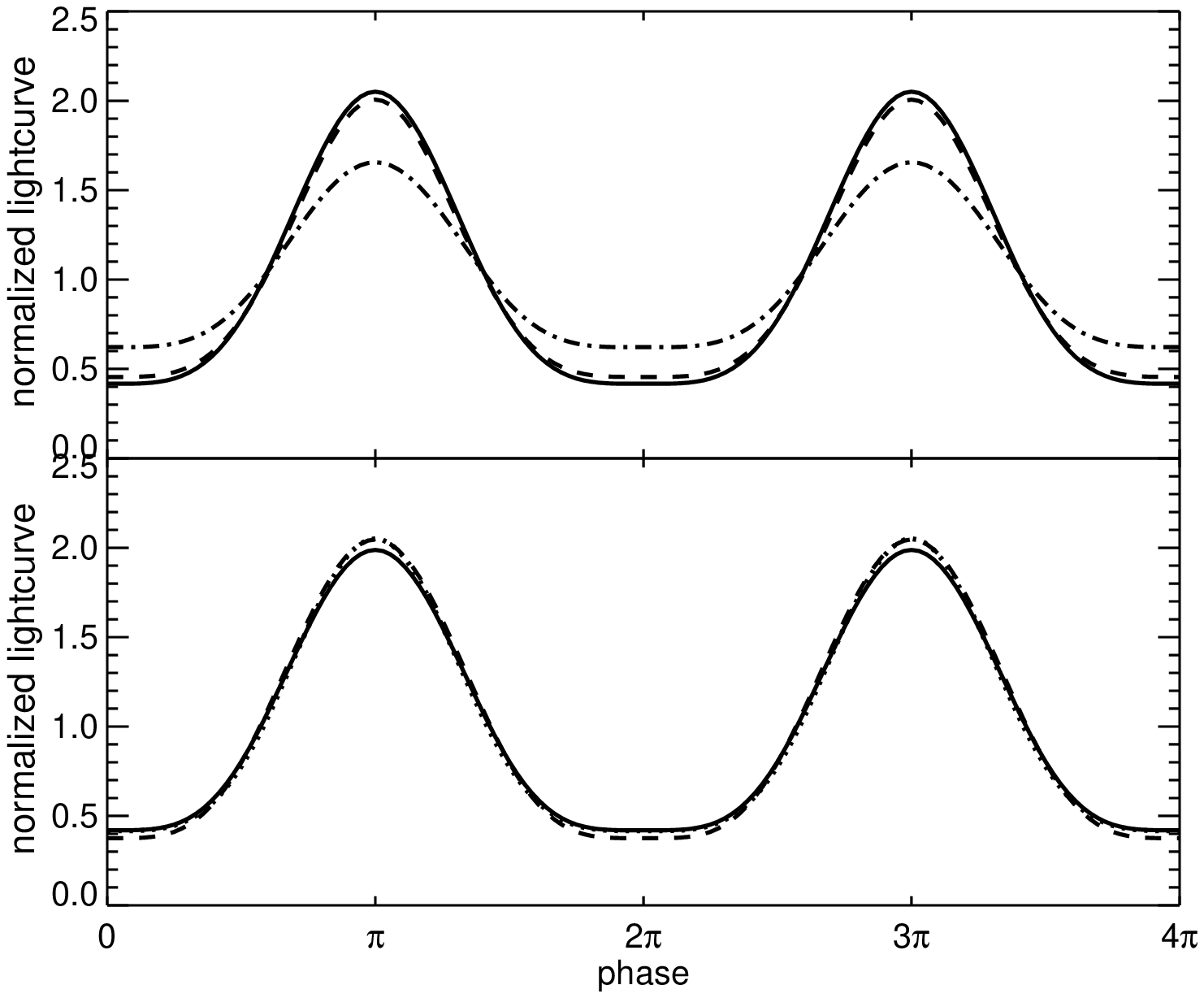}}}
\figcaption{{\it Top panel:} Full-star synthetic light curves at
  photon energies around $1.5$ keV for different crustal magnetic fields. The different
  lines correspond to the different effective temperature
  distributions $T_{\rm eff}(\theta)$ depicted in Figure 3. The pulse
  fractions are 58\% (solid line), 55\% (dashed line) and 38\%
  (dot-dashed line).  {\it Bottom panel:} Light curves for the
  $B_0=2\times10^{14}$ G, $\theta_0=40^\circ$ model at different
  photon energies: $E\sim1.5$~keV (dotted line, pulse fraction 58\%),
  $E\sim3$~keV (solid line, PF 58\%) and $E\sim5$~keV (dashed line, PF
  63\%).}
\end{inlinefigure}

In addition to the crustal toroidal magnetic field, we also include a
star-centered dipole field. The canonical surface dipole field
strength (at the magnetic poles) as inferred from
the $P$-$\dot P$ measurement is 
$B_p\sin\theta_B\simeq 6.2\times 10^{10}$~G (for $R=10$~km and 
moment of inertia $\simeq 10^{45}$~g~cm$^2$), where $\theta_B$ is the angle 
between the magnetic axis and the spin axis. Of course, pulsar spindown power 
is not exactly given by the vacuum dipole radiation formula 
(Contopoulos \& Spitkovsky 2006). 
For concreteness, we adopt $B_p\simeq 8.8\times 10^{10}$~G in our calculations
(corresponding to $\theta_B=45^\circ$), although using a somewhat different value
(within a factor of two) would only have a small effect on our result.
 
Given the magnetic field configuration, the effective temperature
distribution on the NS surface is determined by heat conduction in the
crust (e.g., Schaaf 1990; Heyl \& Hernquist 1998; Potekhin \& Yakovlev 2001)
 Since the magnetic field is symmetric about the magnetic axis,
the effective temperature depends only on the polar angle, $T_{\rm eff} 
= T_\mathrm{eff}(\theta)$.  We use the zero-accretion fitting
formula of Potekhin et al.~(2003) to calculate $T_{\rm eff}(\theta)$.
This formula takes as inputs a core temperature and a crustal magnetic
field of arbitrary strength and direction (assumed constant through
the radial extent of the crust), and returns the effective surface
temperature.  We fix the core temperature at $T_c=10^9$K,
corresponding to $T_{\mathrm{eff}}=3.8\times10^6$~K at the magnetic
poles, which matches sufficiently the hot spot temperature expected
from blackbody fits. 

Figure 3 presents three sample functions $T_\mathrm{eff}(\theta)$, two
at the same value of $B_0$ but different values of $\theta_0$ and one
at a lower value of $B_0$. We also include the corresponding
$f(\theta)$ curves to clarify the effect of changing $\theta_0$.  It
is important to note that when the toroidal field strength $B_0$ is
sufficiently large, even a very smooth $B_\phi$ profile can produce
rather small hot spots.  For example, at $B_0=2\times 10^{14}$~G and
$\theta_0=40^\circ$, the temperature drops from $T_p\simeq 4\times
10^6$~K 

\begin{inlinefigure}
\scalebox{0.55}{\rotatebox{0}{\includegraphics{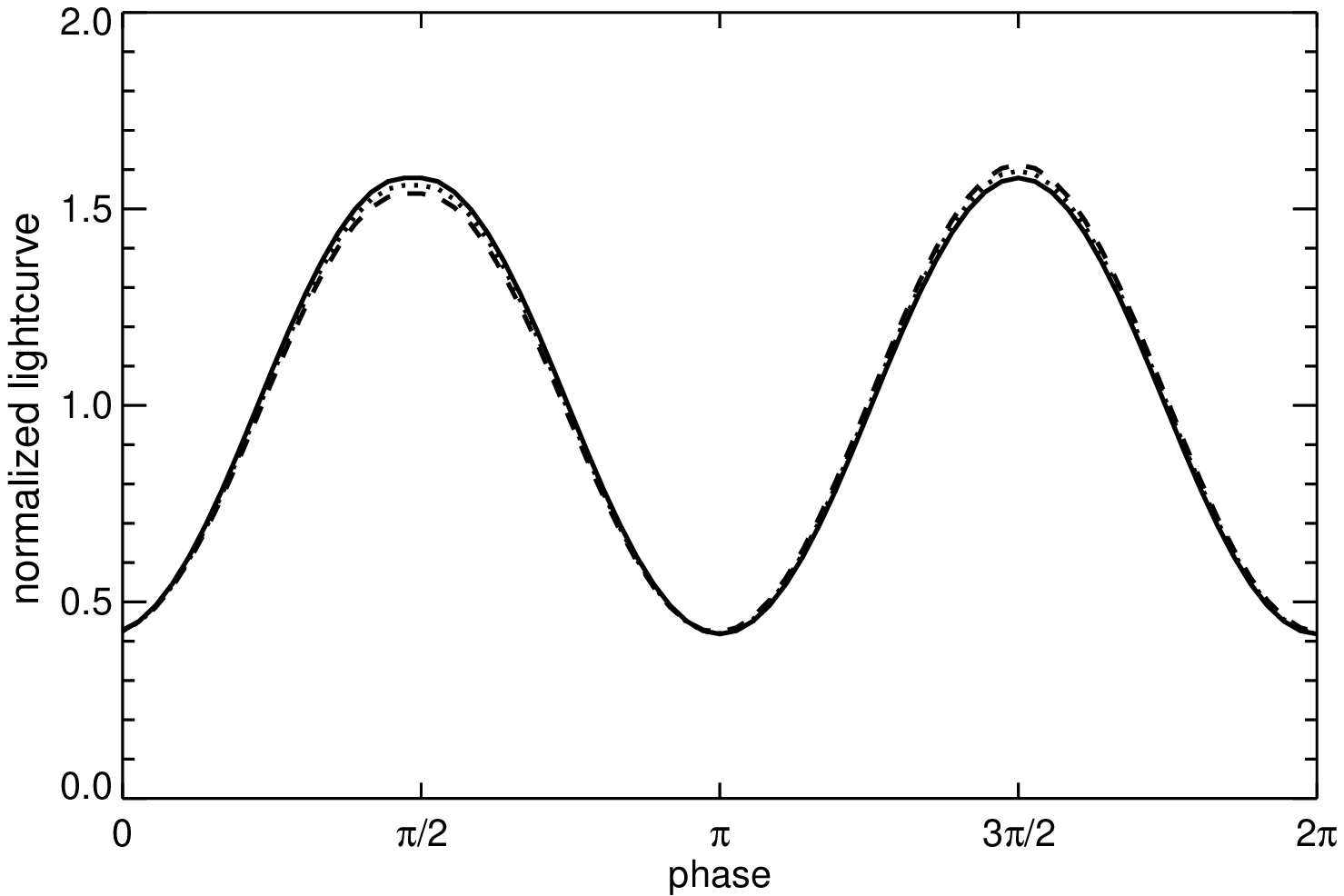}}}
\figcaption{Full-NS light curves at $B_0=2\times10^{14}$ G,
  $\theta_0=40^\circ$, $E=1.5$ keV, and different geometries. Solid --
  $\beta=80^\circ$, $\theta_B=90^\circ$. Dotted -- $\beta=80^\circ$,
  $\theta_B=85^\circ$. Dashed -- $\beta=80^\circ$,
  $\theta_B=80^\circ$.}
\end{inlinefigure}

\noindent at the magnetic pole to $T_p/2$ at $\theta\sim 3^\circ$, and to
$T_p/4$ at $\theta\sim 10^\circ$.

When computing the synthetic X-ray light curves, we split the
effective temperature curve into 7-9 bins, with degree-resolution
close to the pole, and coarser resolution for the rest of the
star. For each bin we create the temperature profile $T(\tau;T_{\rm
  eff})$ at the requisite effective temperature using the TTFT process
described in Section 2. Note that since $B_\phi$ vanishes at the NS
surface, it does not contribute to the atmospheric physics in any way.
We can then compute the light curves for each of the effective
temperature distributions depicted in Figure 3.  

In addition to $B_0$ and $\theta_0$, the light curves also depend on two
geometrical parameters: the angle between the line of sight and the
rotation axis, $\beta$, and the angle between the magnetic dipole axis
and the rotation axis, $\theta_B$. We select $\beta$ and $\theta_B$ in
order to maximize the pulse fraction. Since, as a rule, emission is
strongest for photons emitted parallel to the direction of the
magnetic field, we should select $\beta$ and $\theta_B$ to be the same
or nearly so, such that at some point during the NS's rotation, the
photons we see are emitted parallel to the magnetic field and the
beaming effect is maximized. We also select the angles such that a
single-peaked pulse profile is produced (see below for double-peaked
profiles). After exploring the available phase space, we settle on
$\beta=\theta_B=45^\circ$. This choice is not an overly specific one,
however, as there are many other geometrical configurations that
result in only slightly smaller pulse fractions. Our selection of
$\beta$ and $\theta_B$ maximizes the pulse fraction, and therefore
minimizes how ``dark'' the stellar surface must be away from the hot
spots. This selection is therefore the most conservative estimate of
the required strength of $B_\phi$.

The top panel of Figure 4 presents the whole-star synthetic light curves
at the photon energy $E=1.5$ keV, using each of the three
$T_\mathrm{eff}(\theta)$ functions of Figure 3.  The
$B_0=2\times10^{14}$ G, $\theta_0=40^\circ$ model (solid line)
achieves the required pulse fraction observed in Kes 79 NS.  For
$B_0=2\times10^{14}$ G, $\theta_0=10^\circ$ model (dashed line), the
hot spot is too concentrated with a small area, causing a decrease in
pulse fraction.  For $B_0=2\times10^{13}$ G, $\theta_0=40^\circ$ model
(dot-dashed line), the spot has enough area, but the rest of the star
is simply not dark enough, and thus the pulse fraction is small.  We
therefore conclude that a toroidal magnetic field in the crust of
order $B_\phi\sim{\rm a~few}\times10^{14}$~G or higher, rising slowly
enough from zero toward the magnetic equator, such that it allows the
hot spot to gain enough area, can naturally produce the requisite
pulse fraction observed in the Kes 79 NS.

The bottom panel of Figure 4 shows that the pulse profiles at different
photon energies are approximately the same (with the pulse fraction
slightly higher at $E=5$~keV than at $1.5$~keV), in agreement with the
observation of the Kes 79 NS (Halpern \& Gotthelf 2010). 
Note that while we compute
the pulse profiles at a single well-defined energy, they are nevertheless representative
of the pulse profiles that would be observed in energy bins centered on the quoted energies,
with widths of a few tenths of a keV.

Although our theoretical light curves reproduce the large pulse
fraction observed in the Kes 79 NS, there remains appreciable difference
between the shape of the light curves presented in Figure 4
and the observation. Our model light curves have relatively sharp peaks, with
the system spending nearly half of the rotation period at the
(observed) flux minimum. This is different from the broad flux maximum
observed in Kes 79 NS. 
Note, of course, that we have only considered one particular
functional form for the magnetic field distribution, and the form we
consider is very smooth. It is not clear
that more complex magnetic field configurations would yield the right pulse
shape while maintaining a high pulse fraction.
Another possible way to resolve this difference is to consider pulse
profiles with two peaks per rotation period, one peak for each
magnetic hot spot. This is possible when both the line of sight and
the magnetic dipole axis are nearly perpendicular to the rotation axis.
Such a light curve would be observationally indistinguishable from a
single-peaked light curve, but would imply that the actual rotation
period of the NS is two times shorter.

Figure 5 presents several examples of the two-peaked light curves for
different combinations of the geometric parameters $\beta$ and
$\theta_B$ (both around $90^\circ$). We see that the width of the
peaks is much closer to what is observed in Kes 79 NS.  Obviously,
when either $\beta$ or $\theta_B$ deviates significantly from
$90^\circ$, the two peaks will have unequal heights, which can be
measurable. Still, our result in Figure 5 shows that 
the square of geometrical phase space with corners at
$(\beta,\theta_B)=(80^\circ,80^\circ)$ and (by symmetry)
$(110^\circ,110^\circ)$ provides viable 
two-peaked light curves that may be indistinguishable from 
single-peaked light curves.

\section{Discussion}

\subsection{Large Pulse Fraction - How?}

Beloborodov (2002) demonstrated that two infinitesimal antipodal hot
spots, emitting isotropically, on the surface of a $1.4\mathrm{M}_\mathrm{sun}$, $10$ km NS cannot, in full geometric generality,
produce light curves with pulse fractions higher than about
$12\%$. Poutanen \& Beloborodov (2006) later generalized this to anisotropic
emergent intensities, producing pulse fractions of $\mathrm{a~few}\times 10\%$.
Thus, it seems that 
even if one were to find a way to create hot spots on the NS surface,
it would be difficult produce the 60\% pulse fraction observed in Kes
79 NS. How, then, did we accomplish this?

In fact, diffuse hot spots of finite size (with varying temperatures),
combined with the beaming due to anisotropic photon opacities in
magnetic fields, are essential for generating the large pulse fraction
in our model. In our calculations, the hottest regions with radii of
$\sim 0.5^\circ$ around each magnetic pole do not produce pulse
fractions higher than about $40\%$. Instead, it is the subsequent
annuli of decreasing effective temperature that create the bulk of the
pulsed emission. Therefore, to quantitatively model the light curves of
NS surface emission, it is important to go beyond the single-temperature
hot spot approximation widely used in many previous studies. It is
for this reason that a reasonably efficient method, such as TTFT developed
in Section 2, is necessary in order to add up atmosphere emissions from 
many different patches of the NS surface to produce synthetic light curves.

\subsection{Crustal Magnetic Field Evolutionary Timescale}

The crustal magnetic field assumed in our model (Section 3) 
is unlikely to be unique for producing the observed light
curve of Kes 79 NS. In a way, it is the most conservative
magnetic configuration, with the minimum number of free 
parameters, and allows for self-consistent calculation
of the surface temperature distribution and the resulting
light curve. One can invoke more complex magnetic field
topology, but the general conclusion of our result, that 
the NS in Kes 79 contains a significant hidden magnetic field
(of order a few$\times 10^{14}$G or larger) in the bulk region
of its crust, is likely to be robust. 
We also note that while the ``strong toroidal field + weak dipole
field'' configuration adopted in our model has generally been thought
to be unstable in fluid stars (e.g., Braithwaite \& Nordlund 2006;
Reisenegger 2009), this may not, in fact, be the case, as Braithwaite
(2009) demonstrated that the effect of gravitational stratification is
such that even a relatively weak poloidal field can be enough to
stabilize the configuration. In NSs beyond the fluid stage of
evolution, the crust also has a stabilizing effect.
If we include a poloidal field of similar strength as the toroidal field, 
our general conclusion will be unchanged.

While it is beyond the scope of this paper to speculate 
how the magnetic field configuration
adopted in our model may come about, we can estimate how long this
magnetic field configuration lasts in its current form to allow for a
good chance of being observed. 
Given the toroidal field strength we have found above for Kes 79 ($\sim 10^{14}$~G), its evolution should be dominated by Hall drift, since the Ohmic timescale for a young NS is much longer than the Hall timescale (see Reisenegger et al.~2005; Pons \& Geppert 2007; Pons et al. 2009):
\begin{equation}
\label{eq:halleq}
{\partial\textbf{B}\over\partial t}=-\nabla\times\left[{c\over 4\pi
    n_e e}\left(\nabla\times\textbf{B}\right)\times\textbf{B}\right],
\end{equation}
where $n_e$ is the electron number density. The characteristic Hall 
timescale is 
\begin{equation}
\label{eq:thall}
t_{\rm Hall}={4\pi n_e e L^2\over c B},
\end{equation}
where $L$ and $B$ are the typical length scale and field strength.
Using values $n_e\sim10^{35}\,\mathrm{cm}^{-3}$, $L\sim0.5H$, with $H$ the crust
thickness $\sim1~\mathrm{km}$, and $B\sim2\times10^{14}$~G, we find
$t_{\rm Hall}\sim 8~\mathrm{kyr}$.  We can further evaluate the Hall
timescale as a function of location in the crust via $t_{\rm Hall}\sim
|{\bf B}|/|\partial {\bf B}/\partial t|$. 
We find that the Hall time scale is greater than $10$ kyr throughout
most of the crust, and decreases to a value of $5$ kyr at the edges
of the crust. 
Thus, the magnetic field configuration is stable enough to remain
virtually unchanged up to more than the current age of the NS
(about 7~kyr).

\section{Conclusion}

This paper has two main results: 

(i) We have developed a general-purpose method (``Temperature Template
with Full Transport'', TTFT) to efficiently compute the radiation intensity
from different patches of a NS surface with arbitrary magnetic fields and
effective temperatures. The TTFT method accounts for the effects of
realistic polarized photon opacities and anisotropic radiative
transport in NS atmospheres without self-consistent atmosphere
modeling for every surface patch. This method is therefore useful for
computing synthetic light curves of X-ray emission from the whole
neutron star surface.

(ii) We have used the TTFT method to model the X-ray light curve of
the young NS in the Kes 79 supernova remnant. We find that a crustal magnetic
field of order a few$\times 10^{14}$G, much stronger than the
observed dipole field, can produce significant surface temperature
inhomogeneity and generate the observed large pulse fraction. Small
hot spots can be produced even for fields that vary smoothly from the
magnetic poles to the equator. Although our specific crustal field model
adopts a conservative axisymmetric toroidal configuration,
we believe that our general conclusion about the strength of the hidden
magnetic field in the star's crust is robust for more general field
configurations.

Although we have developed the TTFT method with the application to the Kes
79 NS in mind, the basic methodology may prove useful for wider
applications.  In many published works on surface emission from
magnetic NSs, the observed radiation is compared to a
single-temperature NS atmosphere model, while in reality, a synthetic
radiation flux from the whole star is needed. Using the TTFT
method, the total radiation from different patches of the NS may be
calculated with temperature templates from a small number of
self-consistent atmosphere models. We have shown in this paper that 
at least with respect to the light curves, the TTFT method can produce
accurate result.

Our finding that the Kes 79 NS has a much larger hidden magnetic field
in the crust than the dipole field inferred from the spindown
measurement (see Halpern \& Gotthelf 2010) may have important
implications for the nature of central compact objects (CCOs) and the
evolution of NS magnetic fields. CCOs constitute a significant
population of NSs in supernova remnants. Recent observations have
shown that the 7-10 known CCOs (or candidates) have weak dipole magnetic
fields (Halpern \& Gotthelf 2010,2011), so their relationship to other
types of isolated NSs (radio pulsars, magnetars and dim isolated
NSs) is unclear (van Kerkwijk \& Kaplan 2007; Kaspi 2010). 
The presence of a large sub-surface magnetic field in the Kes 79 NS
supports the notion that supernova fall-back may have submerged the
magnetic field, which may later diffuse back to the surface (e.g.,
Muslimov \& Page 1995; Geppert et al.~1999; Ho 2011).  Together, this
property of the Kes 79 NS and the recent discovery of SGR 0418+5729
with a weak dipole field (Rea et al. 2010; Turolla et al.~2011)
suggest that both the hidden sub-surface magnetic field and the
``visible'' dipole field of NSs play an important role in their
observational manifestations.

\medskip
We thank Wynn Ho and Valery Suleimanov for supplying their neutron star 
atmosphere models, which are essential for calibrating our method.
We also thank Wynn Ho, Alexander Potekhin and Marten van Kerkwijk for useful discussion
and suggestions. This work was supported in part by NSF grant
AST-1008245 and NASA grant NNX10AP19G.

\end{document}